\documentclass[
twocolumn,
]{ceurart}
\PassOptionsToPackage{table}{xcolor}
\usepackage{listings}
\usepackage{hyperref}
\usepackage{booktabs}
\usepackage{xcolor} 
\usepackage{array}
\usepackage{siunitx}
\usepackage{multirow}
\usepackage{threeparttable}
\usepackage{geometry}
\usepackage{float}
\usepackage{pdfpages}
\usepackage{subcaption}
\definecolor{rowgray}{gray}{0.95} 

\begin{document}

\copyrightyear{2023}
\copyrightclause{Copyright for this paper by its authors. Use permitted under Creative Commons License Attribution 4.0 International (CC BY 4.0).}
\conference{SDU@AAAI’24: Workshop on Scientific Document Understanding,
February 26, 2024}

\title{Temporal Graph Neural Network-Powered Paper Recommendation on Dynamic Citation Networks}

\author[1]{Junhao Shen}[
email=junhaos@smu.edu
]
\cormark[1]
\fnmark[1]

\author[1]{Mohammad Ausaf Ali Haqqani}[
email=malihaqqani@smu.edu,
]
\fnmark[1]
\author[1]{Beichen Hu}[
email=beichenh@smu.edu,
]
\author[1]{Cheng Huang}[
email=chenghuang@smu.edu,
]
\author[1]{Xihao Xie}[
email=xihaox@smu.edu,
]
\address[1]{Department of Computer Science, Southern Methodist University, USA}

\author[2]{Tsengdar Lee}[
email=tsengdar.j.lee@nasa.gov,
]
\address[2]{Science Mission Directorate, NASA Headquarters, USA}

\author[1]{Jia Zhang}[
email=jiazhang@smu.edu,
]

\cortext[1]{Corresponding author.}
\fntext[1]{These authors contributed equally.}

\begin{abstract}
Due to the rapid growth of scientific publications, identifying all related reference articles in the literature has become increasingly challenging yet highly demanding. Existing methods primarily assess candidate publications from a static perspective, focusing on the content of articles and their structural information, such as citation relationships. There is a lack of research regarding how to account for the evolving impact among papers on their embeddings. Toward this goal, this paper introduces a temporal dimension to paper recommendation strategies. The core idea is to continuously update a paper's embedding when new citation relationships appear, enhancing its relevance for future recommendations. Whenever a citation relationship is added to the literature upon the publication of a paper, the embeddings of the two related papers are updated through a Temporal Graph Neural Network (TGN). A learnable memory update module based on a  Recurrent Neural Network (RNN) is utilized to study the evolution of the embedding of a  paper in order to predict its reference impact in a future timestamp. Such a TGN-based model learns a pattern of how people's views of the paper may evolve, aiming to guide paper recommendations more precisely. Extensive experiments on an open citation network dataset, including 313,278 articles from \href{https://paperswithcode.com/about}{PaperWithCode}, have demonstrated the effectiveness of the proposed approach. 
\end{abstract}


\maketitle

\section{Introduction}
With the rapid proliferation of scientific publications, it has become increasingly more challenging, yet highly demanding, for researchers to find proper reference papers for their papers under construction. Recent years have witnessed a number of methods aiming for scientific paper recommendation, including content-based filtering, collaborative filtering, co-occurrence, graph-based, global relevance, and hybrid models \cite{kreutz2022scientific}. The advancement of Graph Neural Networks (GNNs) has marked a significant stride in learning representations of graph-structured data, enabling graph-based methods to effectively learn citation relationships among papers. However, most GNN-based models are designed oriented to static graph structures, meaning that they treat each paper citation as a static existence instead of considering its occurrence timestamp.

The intricate interconnections among papers (i.e., nodes in a citation network) are in constant flux, evolving with each new citation. Citations reported at different time frames may represent different users' views of a paper in the community at the time. For example, a paper published in an image processing conference may be first cited by image processing papers; however, several years afterward, people may cite the paper mainly for its innovative machine learning algorithm. Such recent community views shall impact the embedding of the paper with a higher weight. Furthermore, a paper currently searching for reference papers will be published in the future. Thus, it would be ideal if such reference papers were selected in the context of a future time spot. To the best of our knowledge, the consideration of papers' publication timestamps and dynamic citation relationships has been largely overlooked in the existing studies.

In this project, we aim to fill the gap by adding a time dimension in the consideration for paper recommendations. We hypothesize that the impact of a citation event on a paper should be associated with the timestamp when the citation happens. Our core idea is to update the embedding of a paper whenever new papers are published and added to its connected citation network. The collection of the embeddings of the paper over the years, since its publication date, forms a time series and can be used to predict its possibility of being cited by others in the future. In this way, we introduce a time dimension into a citation network, where each citation relationship comes with a timestamp, and nodes embeddings evolve over the time as the new citations are added to the existing graph.
 
To establish a temporal citation network and capture its temporal dynamics, we utilize a Temporal Graph Neural Networks (TGN)-based memory module \cite{Rossi2020TemporalGN} to update node (i.e., paper) embeddings in a continuous-time sequence. Additionally, a learnable message module is built to prevent excessive message interchanges over time. Such a setting enables effective aggregation of continuous-time dynamic interactions within the network.

Meanwhile, we use citation relationship and encoded time feature as edge attributes to perform regularized propagation in the network. A Graph Transformer convolutional (TransConv) layer \cite{Shi2020MaskedLP} is employed, together with a multi-head attention mechanism.

On top of the constructed temporal citation network, we designed and developed a TGN-Transformer-based recommendation (TGN-TRec) engine for paper recommendations. Experiments over an open dataset, which includes machine learning-related papers from \href{https://paperswithcode.com/about}{PaperWithCode}, have demonstrated the effectiveness of our method in terms of paper recommendation accuracy, compared with state-of-the-art approaches.

The contributions of this paper are three-fold. First, we introduce a time dimension into a citation network to carry its temporal dynamics. Such a temporal citation network captures both the temporal structure among papers and evolving embeddings of papers, which paves a new way for people to better understand the academic influence among papers over time. Second, we report a practical engineering methodology to construct a temporal citation network. Third, we present a temporal graph neural network-powered paper recommendation engine.

The remainder of the paper is organized as follows. Section 2 discusses related research work. Section 3 presents the TGN-TRec paper recommendation engine over a temporal citation network. Section 4 presents experiments and discusses empirical results. Section 5 draws conclusions.
\\

\section{Related Work}

This section discusses related work from three aspects: graph neural networks, dynamic graphs for citation networks, and temporal graph neural networks in scientific paper recommendations.

\subsection{Graph Neural Networks}
Graph Neural Networks (GNNs) have revolutionized the way we approach link prediction problems in graphs, by enabling the learning of complex node representations that capture the structural context of each node within a graph \cite{Wu_2021}. Recent advance of  GNN focused on static graphs, learning structure information by performing message passing mechanisms between embeded nodes, the learned node embedding can be used for various prediction tasks. The Graph Convolution Networks(GCN)\cite{kipf2017semisupervised} proposed a fast spectral-based graph convolution kernel that makes GNN efficiently perform on node classification tasks. GraphSAGE\cite{Hamilton2017InductiveRL} proposed a topological-based method by using sampling and aggregating strategy to make GNN can be applied to inductive learning tasks. In Graph Attention Network (GAT)\cite{veličković2018graph}, the researchers applied a multi-head attention mechanism by using learnable matrices for weighting the importance of neighbor nodes, thus optimizing the message passing process. Those methods have been applied to a variety of tasks\cite{gnn_applications_zhou}, ranging from social network analysis to protein-protein interaction analysis, and knowledge graph areas.

\subsection{Dynamic Graphs for Citation Networks}
Unlike static networks, temporal networks are characterized by edges that form or dissolve over time, requiring specialized models that can account for these dynamics. The dynamic graph representation learning can learn dynamic graph that evolves over time or events \cite{Kazemi2020RepresentationLF}. There are two categories of dynamic graphs: discrete-time dynamic graph and  continuous-time dynamic graph. A discrete-time dynamic graph (DTDG) is a sequence of static graph snapshots over time, where the edges in each snapshot of graph take the same timestamp. Discrete-time approaches segment the network into time slices and analyze each slice independently or in sequence, some approaches perform graph learn on graph snapshots by applying static methods\cite{MA2019100490,Holm2020LongitudinalCP}.

A continuous-time dynamic graph (CTDG) represents a dynamic graph whose comprising node pair interactions evolve over time. It illustrates changes in a graph in a more general way. Recent advancements of continuous-time models aim to capture the network's evolution at a finer granularity, applying sequence-based methods to update node information by capturing nodes' interaction sequentially \cite{Kumar_2019, Trivedi2017KnowEvolveDT, cui2021dygcn}. The Temporal Graph Attention Network (TGAT) \cite{Xu2020TGAT} proposes a functional time-encoding module to learn dynamic interactions as a graph evolves. The TGN \cite{Rossi2020TemporalGN} put temporal graph neural networks into a framework by proposing an RNN-based memory update module. These temporal models have been shown to be particularly effective in capturing the causality and sequential dependencies inherent in temporal networks.

\subsection{Temporal Graph Neural Networks in Scientific Document Recommendation}
GNN has been proven its successful application and great potential power of application on recommendation systems \cite{wu2022graph, gao2023survey}. As a subdomain of GNN and application of a recommendation system, citation networks-based recommendations present a unique challenge for link prediction due to their directed nature, the evolution of research topics over time, and the presence of citation lags. Traditional heuristics such as the clustering analysis have been applied to citation networks with limited success \cite{newman2001clustering}. Machine learning approaches, particularly those employing GNNs, have shown improved performance by utilizing not only the content but also the network structure of the papers by message passing mechanism \cite{kreutz2022scientific}. 

To capture the dynamic nature of entities for a continuous-time bipartite graph scenario, researchers applied Temporal Graph Sequential Recommender(TGSRec) \cite{fan2021TGSRec} to capture dynamics collaborative signals from both users and items in a sequential pattern. However, there is limited research on a recommendation method for scientific documentation that considers communities' view of the existing paper evolving with the new citation. Apart from \cite{Holm2020LongitudinalCP} that only predicts citation counts by using GNN on static snapshots of citation networks over years, our model considers edge-level timestamp that is a continuous-time evolving dynamic citation networks. For supporting the scientific paper recommendation system, our model not only predicts citation counts but also potential citation probabilities in future time spans.

\begin{figure}
  \centering
  \includegraphics{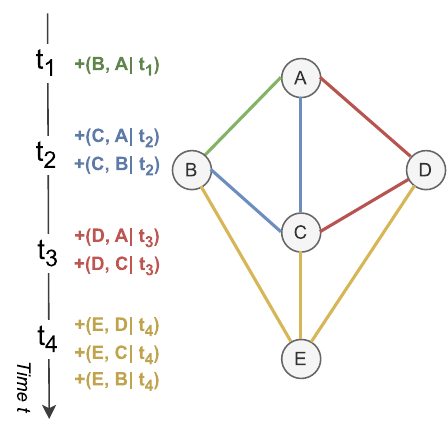}
  \caption{Illustration of dynamic citation networks, the graph will incrementally expand as time.}
  \label{fig:dy_cite_network}
\end{figure}

\section{Methodology}
\begin{figure*}
  \centering
  \includegraphics[width=\textwidth]{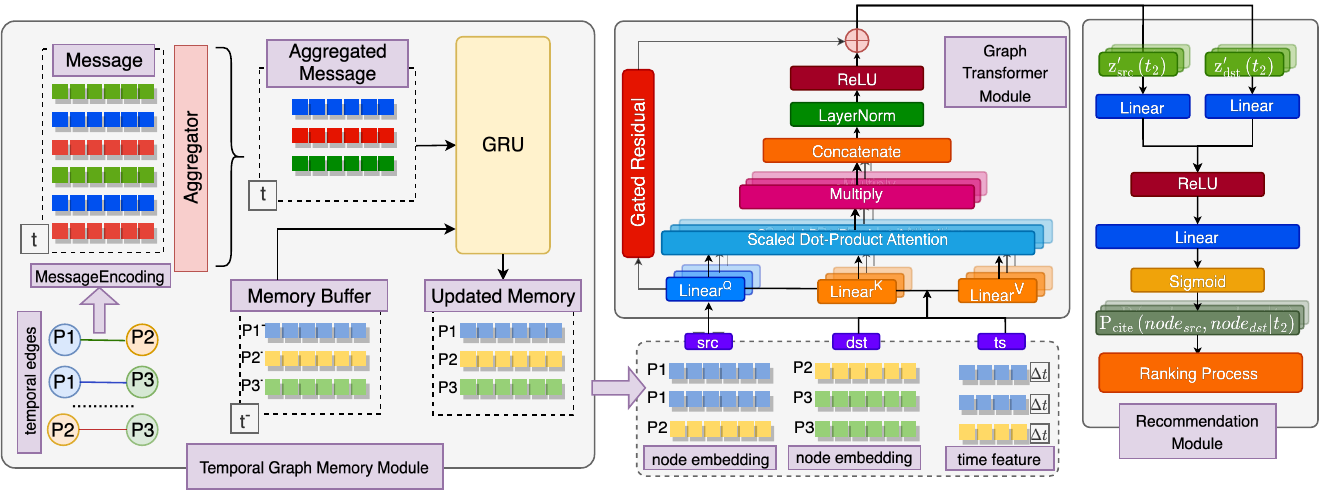}
  \caption{The illustration of the model processing batch of interactions between nodes $P1$, $P2$ and $P3$, the temporal graph module computes messages for each interaction and using an aggregate function to merge messages that send to each node, the GRU cell will take aggregated messages and previous state/meomory of each node and output is nodes' state/memory in the latest timestamp. For performing transformer convolution operation in the Graph Transformer Module, we use source node state, destination node state and edge attribute(encoded time different) as Q, K, V to a scaled dot-product attention layer. The output node embeddings of the graph transformer module are used for computing citation scores in the Recommendation Module.}
  \label{fig:tgn_citation_rec}
\end{figure*}

We introduce our paper recommendation system based on the Temporal Graph Neural Network(shown as Figure \ref{fig:tgn_citation_rec}). The major components of the model contain a Temporal Neural Network (TGN) Memory Module \cite{Rossi2020TemporalGN} as an encoder to learn paper citation relationships in a dynamic way, and an attention-based prediction module for paper recommendation as a decoder. We further implemented the TGN-based encoder, by setting a self-learnable message module to adaptively compute messages between nodes in order to prevent excessive messaging passing in an evolving graph.

\subsection{Static Graph Representation Learning}
In a static graph $\mathcal{G}=(\mathcal{V}, \mathcal{E})$, the node set $\mathcal{V} = \left\{v_1, v_2, ...v_N \right\}$, and $N$ is the number of nodes; $\mathcal{E}$ denotes a collection of edges $e_{ij}$, where $e_{ij} = (v_i, v_j)$ for all $i, j = 1, 2, ... N$. In a graph neural networks scenario, we usually have a node features $\mathcal{X} = \left\{x_1, x_2, ...x_N \right\}$. The topological method of graph neural networks usually uses a message-passing framework that creates hidden nodes embedding $\mathbf{z}_i$ by aggregating neighbor nodes' information in the form:
$$
\mathbf{z}_i=\gamma_{\Theta}\left(\mathbf{x}_i, \bigoplus_{j \in \mathcal{n}(i)} \mathbf{m}_{i j} \right),  \quad\quad \mathbf{m}_{i j}=\phi_{\Theta}\left(\mathbf{x}_i, \mathbf{x}_j, \mathbf{e}_{j, i}\right),
$$
where $e_{ij}$ is edge features, $m_{ij}$ is message computed by a message module, where $\bigoplus$ denotes a differentiable, permutation invariant function, e.g., sum, mean, min, max or multiply, $\gamma_{\Theta}$ and $\phi_{\Theta}$ denote learnable functions such as linear or attentional layer. $\mathcal{n}(i)$ denotes neighbors node for node $v_i$. 

\subsection{Dynamic Graph Representation Learning}
In a Dynamic Citation Network(shown as Figure\ref{fig:dy_cite_network}), the node interactions are a sequence of citation relationships between papers, for each edge (paper $v_i$ cite paper $v_j$) $e_{ij}(t)$ have a timestamp $t$, since citation networks only have addition operation, and new citation relations happen when a new node is added to the graph. We also consider the influence transferring of an existing paper in the citation network, so the message passing is bidirectional, the existing node's embedding changes when a new node is added into the graph. The temporal graph can be denoted as $\mathcal{G(\mathrm{T})}=(\mathcal{V(\mathrm{T})}, \mathcal{E(\mathrm{T})})$, $\mathcal{G(t)}$ represents a temporal citation graph at timestamp $t$ where $t \in \mathrm{T}$. Thus the hidden node embedding at timestamp $t$ is $\mathbf{X(t)} = \left\{x_1(t), x_2(t), ...x_N(t) \right\}$.

\subsection{Memory Module}
To capture long-term memory when a new node has been added to the graph, we adopted the memory module proposed in TGN \cite{Rossi2020TemporalGN}. The existing papers in the temporal citation graph will update their memory when new paper cite them, this module also allows the existing papers to keep their original features and interaction history with other papers in a compress format. Different from the implementation in TGN, our model takes papers' text embedding from SciBert as their initial state $S(t_0)$ when they are added to the citation graph. It will aggregate the messages from their neighbor papers and update the memory when new papers cite them over time. We use the same annotation to represent the memory module in our model. In the memory module, we have a memory updater that is a recurrent neural network cell for updating the papers' embeddings in a sequential manner. This module can save the initial memory from the paper's abstract and a historical interaction among papers along with the time evolving. In our model, we use GRU \cite{cho2014learning} as the updater, and it takes aggregated information from the paper to cite events on timestamp $t$, the memory update format shows as follows:
$$
\begin{aligned}
& r=\sigma\left(W_{i r} m_i(t)+b_{i r}+W_{h r} s_i(t^{-})+b_{h r}\right) \\
& z=\sigma\left(W_{i z} m_i(t)+b_{i z}+W_{h z} s_i(t^{-})+b_{h z}\right) \\
& n=\tanh \left(W_{i n} m_i(t)+b_{i n}+r *\left(W_{h n} s_i(t^{-})+b_{h n}\right)\right) \\
& s_i(t)=(1-z) * n+z * s_i(t^{-})
\end{aligned}
$$

where $s_i(t)$ is the update state of node $i$ in memory, $s_i(t^{-})$ is the previous state of node $i$ before receiving the aggregate message $m_i(t)$ from its new nodes interaction on time $t$.

\subsection{Message Module}
\subsubsection{Message Encoding}
The messages are computed from every interaction event between new publication papers and existing papers, for considering the impact transferring of a paper, we use bi-directional message passing and the message is computed by following rules:

\begin{align}
\mathbf{m}_i(t) &= \operatorname{msg}_{\mathrm{s}}\left(\mathbf{s}_i\left(t^{-}\right), \mathbf{s}_j\left(t^{-}\right), \Delta t \right), \\
\mathbf{m}_j(t) &= \operatorname{msg}_{\mathrm{d}}\left(\mathbf{s}_j\left(t^{-}\right), \mathbf{s}_i\left(t^{-}\right), \Delta t \right)
\end{align}

where $\mathbf{m}_i(t)$ represents the  message that will be sent from node $i$ to node $j$ and verse versa. We are referring the implementation in TGN which concatenates state of node $i$($\mathbf{s}_i\left(t^{-}\right)$), node $j$($\mathbf{s}_j\left(t^{-}\right)$) in last timestamp and $\Delta t$ and the encoded time difference between the current timestamp $t$ and last timestamp $t^{-}$. $\operatorname{msg}$ is the message encoding module, which can be directly concatenated or processed by a self-learned linear layer, which we will discuss in the experiment.

\subsubsection{Message Aggregator}
We follow the same aggregation mechanism defined in \cite{Rossi2020TemporalGN} to aggregate messages to a given node$i$. In our implementation, we compare the mean aggregator and last aggregator(keep the most recent message in each batch). 
$$
\overline{\mathbf{m}}_i(t)=\operatorname{agg}\left(\mathbf{m}_i\left(t_1\right), \ldots, \mathbf{m}_i\left(t_b\right)\right)
$$
where $t_1, ..., t_b <= t_N$, $t_N$ is the latest timestamp in each batch's interaction.



\subsection{Graph Transformer Module}
Once the memory/state of each paper node is updated, we employ a Graph Transformer Convolution module \cite{Shi2020MaskedLP} to calculate the embedding of the newly added node positioned between nodes $i$ and node $j$ using an attention mechanism. Each citation relation (i.e., source node cites destination node at a timestamp) serves as a training case to be fed into the transformer, simulating a recommendation process. The embedding of the source node serves as a query (Q), while the embedding of the destination node, along with the timestamp features, acts as key (K) and value (V) inputs for optimally fitting the scaled dot-product attention operator. Such a setup updates node embeddings as follows:
$$
\mathbf{s}_i^{\prime}(t)=\mathbf{W}_1 \mathbf{s}_i(t)+\sum_{j \in \mathcal{N}(i)} \alpha_{i, j}\left(\mathbf{W}_2 \mathbf{s}_j(t)+\mathbf{W}_6 \phi\left(t-t^{-}\right)\right)
$$
The attention coefficients $a_{i, j}$ is computed as follows:
$$
\alpha_{i, j}=\operatorname{softmax}\left(\frac{\left(\mathbf{W}_3 \mathbf{s}_i(t)\right)^{\top}\left(\mathbf{W}_4 \mathbf{s}_j(t)+\mathbf{W}_6 \phi\left(t-t^{-}\right)\right)}{\sqrt{d}}\right)
$$
where $\phi(\cdot)$ represents a generic time encoding function, and $d$ is the hidden size of each head.

\subsection{Recommendation Module}
As shown in Figure 2 on the right-hand side, we build a paper recommendation module. In scientific paper recommendation scenario, we assume given a number of pairs of positive node pairs and negative node pairs, based on the embeddings generated from previous Temporal GNN based model, the prediciton module can clearly identify the correct citation and noise information(negative edges) efficiently. We compute the edges scores by using two linear layers to learn the embedding of the source node and the embedding of the destination node at $t$ and combine the output to another linear layer. The feedforward network function is shown as follows:
$$
    Score_{e_{ij}} = \mathbf{W}_{out}(\mathbf{RELU}( \mathbf{W}_i(\mathbf{s_i(t)}) \parallel \mathbf{W}_j(\mathbf{s_j(t)}) ))
$$
where $\mathbf{W}_i, \mathbf{W}_j$ and $\mathbf{W}_{out}$ are Linear layers, $\mathbf{s_i(t)}$ and $\mathbf{s_j(t)}$ are source node embedding and destination node embedding from GNN at time $t$.

\section{Experiments}
In this section, we present the experiments we conducted and the result analysis.

\subsection{Dataset}
The dataset used for this research is from the well-known machine learning community \href{https://paperswithcode.com/about}{PaperWithCode}\footnote{https://paperswithcode.com}. The Papers with Code community focuses on creating a platform that associates machine learning papers, code, and datasets. It covers the latest machine learning-related papers in fields including Computer Science, Physics, Astronomy, Mathematics, and Statistics. To build the citation networks, we retrieved the reference lists of the papers by querying each paper's ArXiv ID from the SemanticScholar API\cite{ammar-etal-2018-construction}. After a filtering process, our dataset for the citation networks includes 313,278 articles from 1900 to 2023 (show as in Table \ref{tab:papers_count_by_year}). The citation network contains 2,233,780 edges from 1985 to 2023. We use the number of days between the citing paper's publication date and the earliest paper's publication date as a basis to compute the edges' timestamps. The edges are sorted by timestamp, allowing the model to train the dynamic citation networks sequentially as the citation relationships are established. We utilized the abstracts and titles of all papers to generate the initial embeddings using SciBERT \cite{beltagy-etal-2019-scibert}, which were employed as node features in our experiments.

\begin{table*}[htbp]
\centering
\caption{Dataset Statistics by Year}
\label{tab:papers_count_by_year}
\begin{tabular}{lrrrrrr}
\toprule
\multirow{2}{*}{\textbf{Year}} & \multirow{2}{*}{\textbf{Number of Papers}} & \multicolumn{5}{c}{\textbf{Reference Count}} \\
\cline{3-7}
& & Total & Mean & Median & Min & Max \\
\midrule
<=2010 & 1,096 & 37,235 & 33.973540 & 25 & 0 & 539 \\
2011 & 374 & 14,256 & 38.117647 & 32 & 0 & 231 \\
2012 & 819 & 33,290 & 40.647131 & 34 & 0 & 326 \\
2013 & 3,438 & 115,239 & 33.519197 & 28 & 0 & 434 \\
2014 & 5,087 & 186,356 & 36.633772 & 30 & 0 & 992 \\
2015 & 8,385 & 326,163 & 38.898390 & 33 & 0 & 691 \\
2016 & 12,008 & 472,393 & 39.339857 & 33 & 0 & 645 \\
2017 & 16,715 & 649,326 & 38.846904 & 33 & 0 & 2,644 \\
2018 & 26,399 & 1,040,789 & 39.425319 & 34 & 0 & 1,613 \\
2019 & 36,890 & 1,500,774 & 40.682407 & 36 & 0 & 1,149 \\
2020 & 50,401 & 2,250,294 & 44.647805 & 39 & 0 & 1,576 \\
2021 & 56,821 & 2,619,295 & 46.097306 & 41 & 0 & 1,086 \\
2022 & 61,783 & 2,909,810 & 47.097260 & 42 & 0 & 696 \\
2023 & 33,062 & 1,569,814 & 47.480915 & 43 & 0 & 772 \\
\bottomrule
\end{tabular}
\end{table*}

\subsection{Evaluation Metrics}
To assess the performance of our TGN-TRec model for scientific paper recommendation, we employed three evaluation metrics: Mean Reciprocal Rank (MRR), Precision@K and Recall@K. Below, we detail each of these metrics and explain their relevance in the context of our model evaluation.

    Mean Reciprocal Rank (MRR):
     We use MRR to evaluate the process that computes scores for a list of positive edges and negative edges, ordered by the probability of correctness. The metric is defined as:
    \begin{equation}
    \operatorname{MRR}=\frac{1}{|Q|} \sum_{i=1}^{|Q|} \frac{1}{\operatorname{rank}_i}
    \end{equation}
    where $\operatorname{rank}_i$ defines the rank of the positive edge in a given list of candidate edges. $Q$ is the total number of queries (edges) for a given source node.

    Precision@K and Recall@K help to assess the effectiveness of a model in predicting a set of papers candidates by giving a query paper,
    where 'K' in refers to the number of top recommendations considered in the evaluation. Precision@K and Recall@K are defined as:
\begin{equation}
    \text{Precision}@K = \frac{\text{Number of Relevant Items in Top K}}{K}
\end{equation}
\begin{equation}
    \text{Recall}@K = \frac{\text{Number of Relevant Items in Top K}}{\text{Total Number of Relevant Items}}
\end{equation}

\subsection{Baselines}
In our experimental setup, we compared our TGN-TRec recommendation model under various settings with three leading static graph models: GraphSAGE \cite{Hamilton2017InductiveRL}, GAT \cite{veličković2018graph} and GIN\cite{xu2019powerful}. To ensure fairness, we created equivalent-sized snapshots for training, validation, and testing across all models by setting all papers before 2021 as training data, 2021-2022 as validation data, and 2022-March 2023 as testing data.
We also explored different configurations of our TGN-TRec model as model variants for paper recommendation.

Message Modules: We assessed the impact of using a simple Identity Message Module (as per TGN's vanilla implementation) against a more complex, self-learned Message Module.

Memory Initialization: We compared memory initialization using semantic information from paper abstracts and titles (via SciBERT\cite{beltagy-etal-2019-scibert}) against a structure-only approach.

Aggregator: We assessed the model with different aggregation approaches, where mean stands for average messages for each node and last stands for only keeping the latest message for aggregation in each batch.

\subsection{Experimental Results}

This section presents the evaluation of our TGN-TRec models in different configurations with different state-of-the-art baseline models by applying them to the task of scientific paper recommendation. The effectiveness of each model is assessed based on its performance in several metrics, including Mean Reciprocal Rank (MRR), Recall, and Precision at various cutoffs.

\subsubsection{Quantitative Evaluation}

We conducted extensive experiments on the dataset to compare the performance of our TGN-TRec models against traditional static graph models like GAT and GIN. The evaluation metrics used for this comparison are MRR, Recall, and Precision, which are pivotal for assessing the recommendation quality in scientific literature.

\begin{table*}[htbp]
  \centering
  \begin{threeparttable}
    \caption{Experiment Results}
    \label{tab:experiment_results}
    \begin{tabular}{
      @{} 
      l
      c
      c 
      c 
      S[table-format=1.4] 
      *{3}{S[table-format=1.3]} 
      *{3}{S[table-format=1.3]} 
      @{}
    }
      \toprule
      \textbf{Encoder} & \textbf{initialization} & \textbf{Message}\tnote{1} & \textbf{Aggregator} & {\textbf{MRR}} & \multicolumn{3}{c}{\textbf{Recall}\tnote{2}} & \multicolumn{3}{c}{\textbf{Precision}\tnote{3}} \\
      \cmidrule(lr){6-8} \cmidrule(lr){9-11}
      & & & & & {\textbf{@10}} & {\textbf{@20}} & {\textbf{@50}} & {\textbf{@10}} & {\textbf{@20}} & {\textbf{@50}} \\
      \midrule
      \rowcolor{rowgray}
      GAT & yes & {N/A} & self-attention & 0.952 & 0.442 & 0.630 & 0.891 & 0.902 & 0.817 & 0.622 \\
      SAGE & yes & {N/A} & mean & 0.960 & 0.442 & 0.631 & 0.891 & 0.900 & 0.817 & 0.623 \\
      GIN & yes & {N/A} & {N/A} & 0.970 & 0.447 & 0.637 & 0.893 & 0.900 & 0.813 & 0.617 \\
      \rowcolor{rowgray}
      TGN-TRec & no & Id\tnote{4} & mean & 0.9375 & 0.430 & 0.620 & 0.881 & 0.890 & 0.800 & 0.600 \\
      TGN-TRec & no & Sl\tnote{5} & mean & 0.7817 & 0.445 & 0.635 & 0.871 & 0.902 & 0.820 & 0.620 \\
      \rowcolor{rowgray}
      TGN-TRec & no & Id\tnote{4} & last & 0.9384 & 0.440 & 0.631 & 0.891 & 0.901 & 0.817 & 0.615 \\
      TGN-TRec & no & Sl\tnote{5} & last & 0.7717 & 0.442 & 0.631 & 0.891 & 0.906 & 0.812 & 0.610 \\
      \rowcolor{rowgray}
      TGN-TRec & yes & Id\tnote{4} & mean & 0.965 & 0.442 & 0.631 & 0.891 & 0.902 & 0.817 & 0.622 \\
      TGN-TRec & yes & Sl\tnote{5} & mean & 0.960 & 0.440 & 0.620 & 0.881 & 0.902 & 0.817 & 0.622 \\
      \rowcolor{rowgray}
      TGN-TRec & yes & Id\tnote{4} & last & 0.970 & 0.450 & 0.680 & 0.920 & 0.921 & 0.831 & 0.641 \\
      TGN-TRec & yes & Sl\tnote{5} & last & \textbf{0.975} & \textbf{0.460} & \textbf{0.690} & \textbf{0.940} & \textbf{0.925} & \textbf{0.835} & \textbf{0.645} \\
      \bottomrule
    \end{tabular}
    \begin{tablenotes}
      \item[1] Message encoding technique used in the model.
      \item[2] Recall at different cutoffs.
      \item[3] Precision at different cutoffs.
      \item[4] "Id" stands for Identity.
      \item[5] "Sl" stands for Self-learned.
    \end{tablenotes}
  \end{threeparttable}
\end{table*}

The results tabulated in Table \ref{tab:experiment_results} provide a comprehensive comparison of the models' performance across various metrics. Notably, the TGN-TRec models with initialized memory exhibit superior Mean Reciprocal Rank (MRR), suggesting their enhanced ability to prioritize relevant documents. The precision metrics further validate the models' effectiveness, with the TGN-TRec variants maintaining high accuracy in the top K recommendations.

\subsubsection{Training Dynamics}

\begin{figure*}
    \centering
    \begin{subfigure}[b]{0.48\linewidth}
        \includegraphics[width=\linewidth]{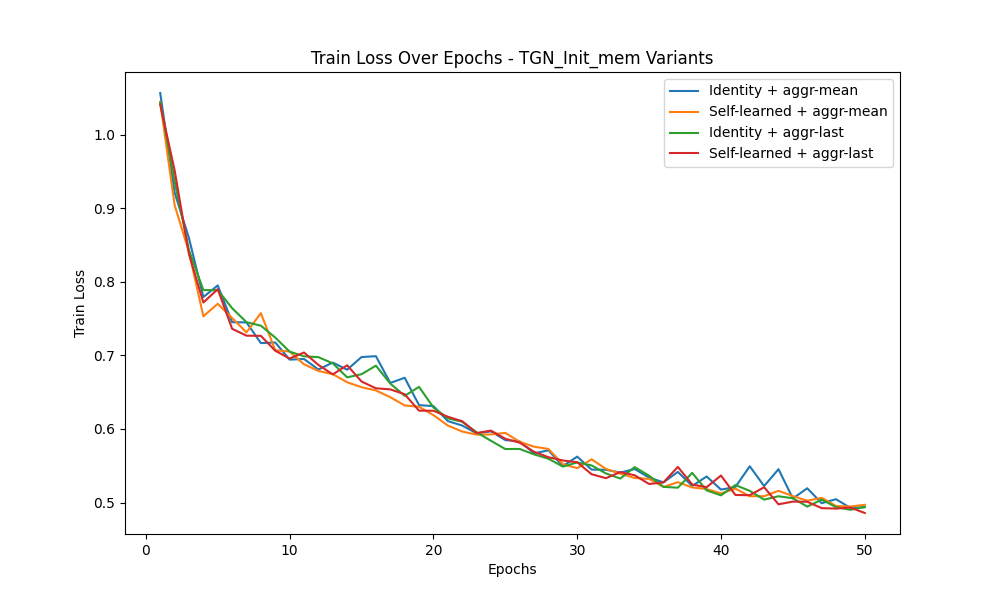}
        \caption{Train Loss for TGN-TRec Init Memory Variants}
        \label{fig:train_loss_tgn_init_mem}
    \end{subfigure}
    \hfill
    \begin{subfigure}[b]{0.48\linewidth}
        \includegraphics[width=\linewidth]{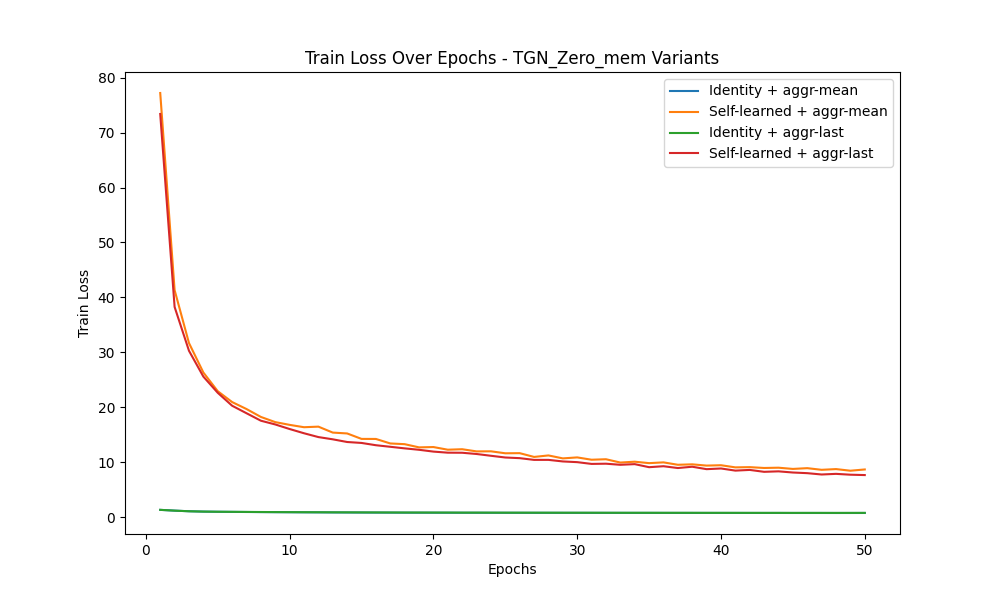}
        \caption{Train Loss for TGN-TRec Zero Memory Variants}
        \label{fig:train_loss_tgn_zero_mem}
    \end{subfigure}
    \caption{Training Loss evolution over epochs for the TGN-TRec models with initialized and zero-initialized memory.}
    \label{fig:training_loss_tgn_models}
\end{figure*}

The training dynamics of TGN-TRec models offer a window into the learning effectiveness of these systems, by visualizing the loss function's decline across epochs, as depicted in Figure \ref{fig:training_loss_tgn_models}. This visualization helps to identify potential overfitting or underfitting, and whether the learning rate is appropriately tuned. A smooth, consistent decline indicates a well-tuned model making steady progress toward optimization.

A critical aspect of the TGN-TRec models' training dynamics is the role of initialization, particularly the use of SciBERT embeddings. Initialization with these embeddings appears to provide a head start to the model by leveraging pre-learned contextual representations, as reflected in the early epochs' rapid loss reduction. This suggests that the model can efficiently abstract higher-level features from the data without needing to learn from scratch, thus potentially reducing training time and resource consumption.
\begin{figure*}
    \centering
    \begin{subfigure}[b]{0.48\linewidth}
        \includegraphics[width=\linewidth]{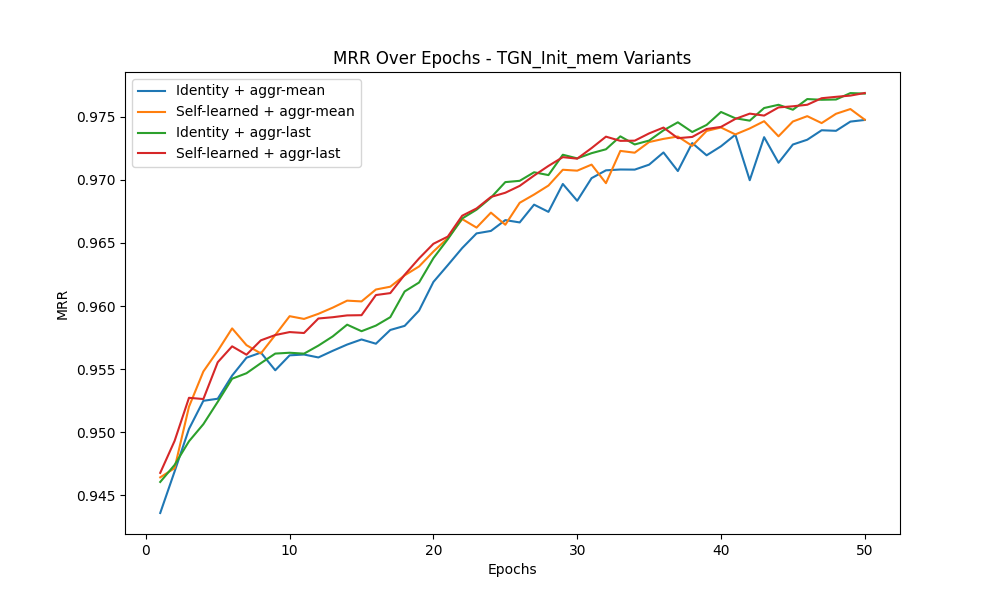}
        \caption{MRR for TGN-TRec Init Memory Variants}
        \label{fig:mrr_tgn_init_mem}
    \end{subfigure}
    \hfill
    \begin{subfigure}[b]{0.48\linewidth}
        \includegraphics[width=\linewidth]{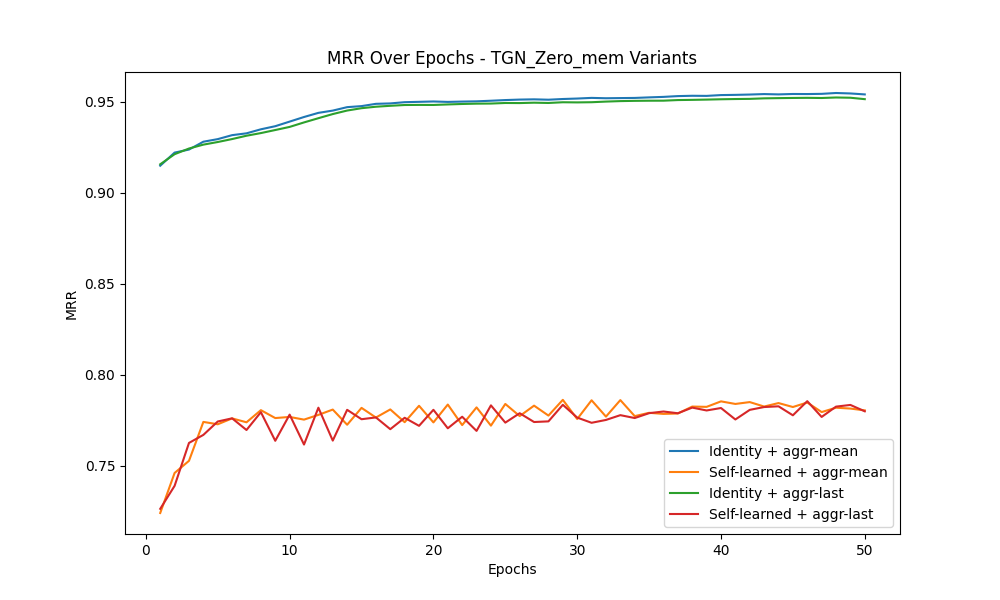}
        \caption{MRR for TGN-TRec Zero Memory Variants}
        \label{fig:mrr_tgn_zero_mem}
    \end{subfigure}
    \vspace{1mm}
    \begin{subfigure}[b]{0.48\linewidth}
        \includegraphics[width=\linewidth]{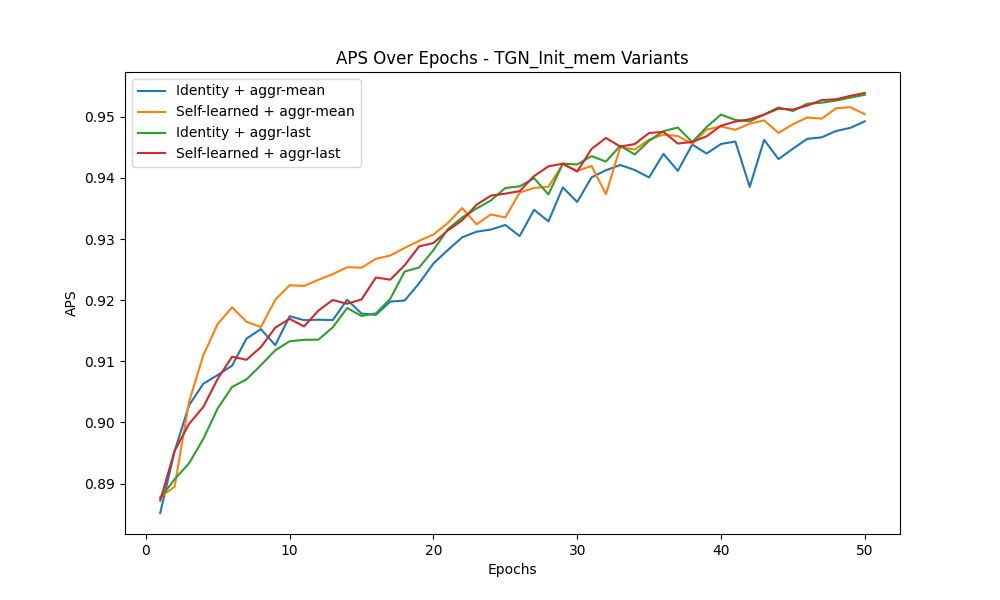}
        \caption{APS for TGN-TRec Init Memory Variants}
        \label{fig:aps_tgn_init_mem}
    \end{subfigure}
    \hfill
    \begin{subfigure}[b]{0.48\linewidth}
        \includegraphics[width=\linewidth]{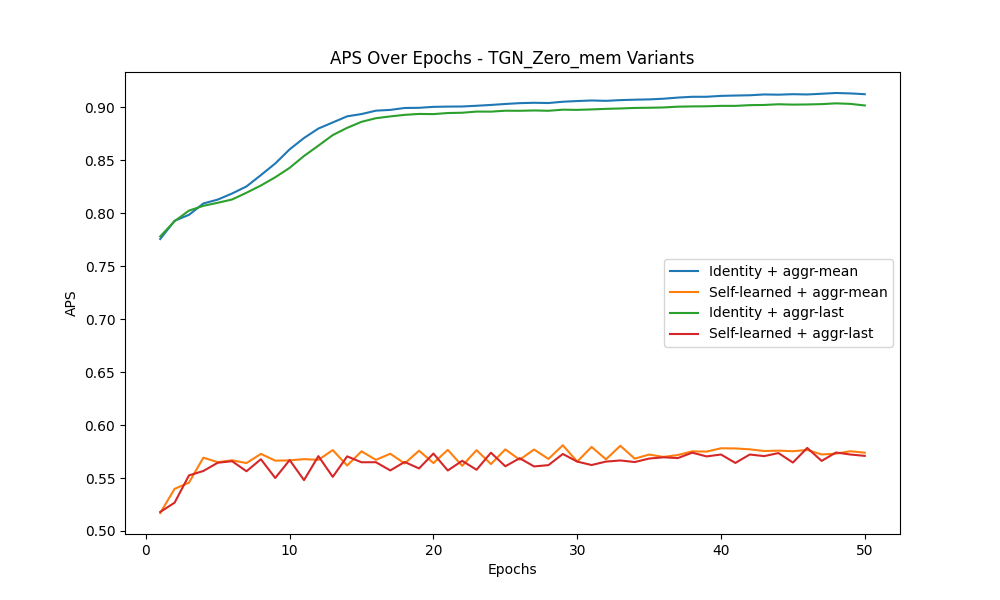}
        \caption{APS for TGN-TRec Zero Memory Variants}
        \label{fig:aps_tgn_zero_mem}
    \end{subfigure}
    \vspace{1mm}
    \begin{subfigure}[b]{0.48\linewidth}
        \includegraphics[width=\linewidth]{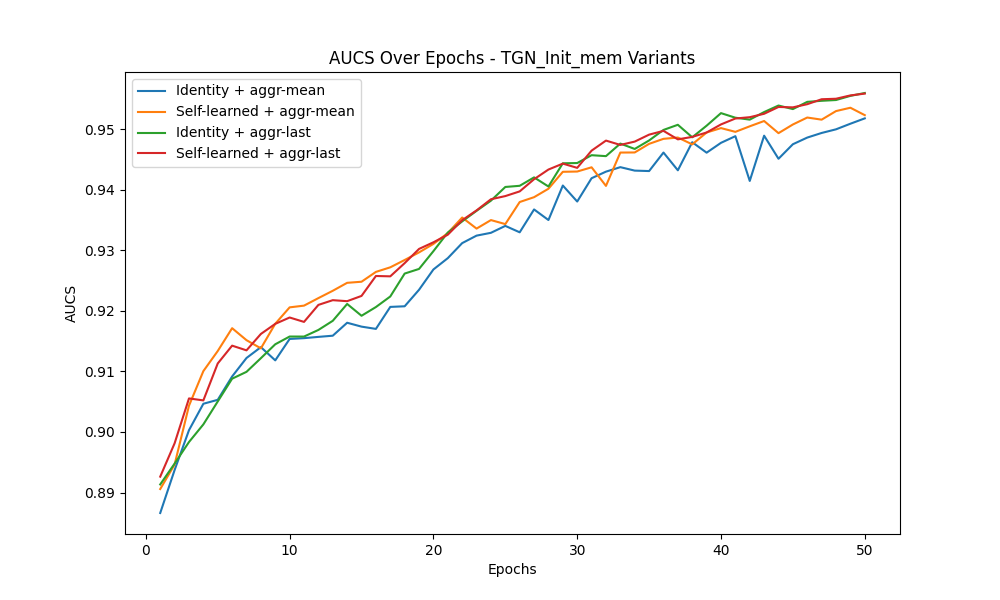}
        \caption{AUCS for TGN-TRec Init Memory Variants}
        \label{fig:aucs_tgn_init_mem}
    \end{subfigure}
    \hfill
    \begin{subfigure}[b]{0.48\linewidth}
        \includegraphics[width=\linewidth]{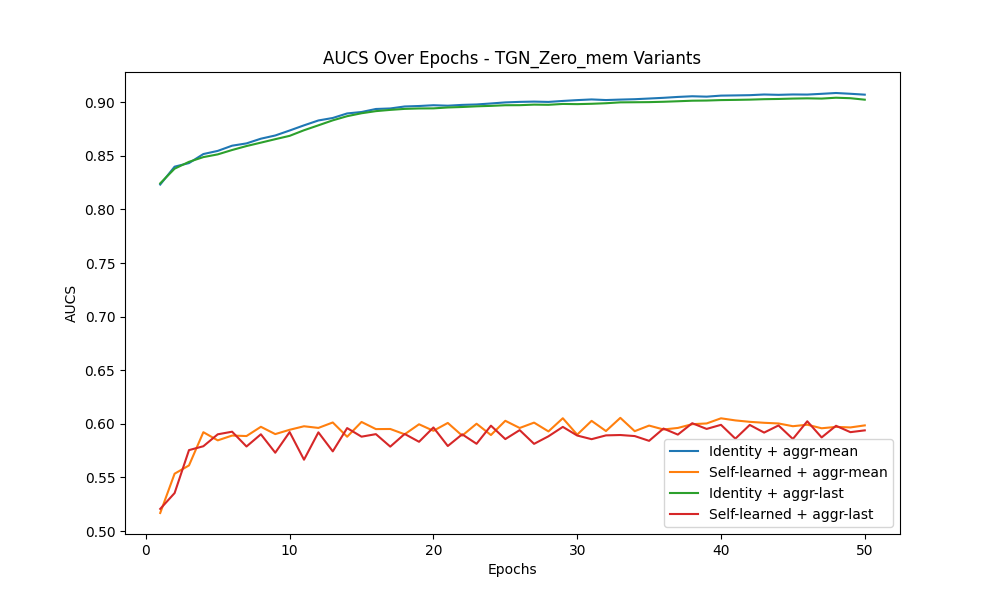}
        \caption{AUCS for TGN-TRec Zero Memory Variants}
        \label{fig:aucs_tgn_zero_mem}
    \end{subfigure}
    \caption{Validation MRR, APS, and AUCS evolution over epochs for the TGN-TRec models with initialized and zero-initialized memory.}
    \label{fig:validation_metrics_tgn_models}
\end{figure*}

In this context, the TGN-TRec models' validation performance, as shown in Figure \ref{fig:validation_metrics_tgn_models}, encompasses several key metrics, including Mean Reciprocal Rank (MRR), Average Precision Score (APS), and Area Under the Curve Score (AUCS). These metrics collectively provide a multi-faceted view of the model's predictive power, robustness against overfitting, and its overall reliability in ranking and recommendation tasks. The TGN-TRec models, through their training dynamics, exhibit signs of such robustness. The consistency in performance metrics across epochs, particularly in scenarios with initialized memories, suggests that the model is learning a stable representation of the data that can withstand the variability inherent in real-world applications.

Finally, the training dynamics also shed light on the computational complexity of the TGN-TRec models. The rate of loss function decline provides indirect evidence of the model's efficiency. A steep initial decrease followed by a plateau suggests that the model quickly captures the primary structure of the data but then requires more nuanced adjustments to refine its understanding. This can influence decisions around early stopping and computational resource allocation, ensuring that the model remains both effective and efficient.

\subsection{Discussions}
Based on the comparative analysis of models under different configurations, we can easily find
static GNN models such as GAT \cite{veličković2018graph} and GIN \cite{xu2019powerful}, provide the same performance as TGN-TRecs without text-based embedding initialization, Since these baseline models were originally designed for static data, they easily fall into overfitting. In addition, the MRR evaluation here only considers the rank of a positive candidate and negative candidate pair, so the MRR score is relative less representative for evaluating model performance in a rigorous scenario. In Precision@K and Recall@K, which showed a comprehensive performance of the models, we strictly ran the models in different negative sampling strategies by setting K in 10, 20 and 50. We found the TGN-based models can easily capture the dynamics of the citation networks, meanwhile the self-learned message module with the last aggregator that keep the most updated message has best performance in overall cases.

\section{Conclusions and Future Work}

In this paper, we introduced the concept of temporal graph into citation work. A time dimension is added, allowing us to predict the future impact of scientific papers based on how their influence evolves over time. The use of continuous-time dynamic graph representation learning allows for a more granular understanding of how a paper's influence develops and changes over time. As a proof of concept, we reported an implementation based on Temporal Graph Neural Networks (TGNNs) and a memory update mechanism based on recurrent neural networks. 

We plan to continue our research work in the following four directions. First, we will study the scalability of our model over large-scale datasets, especially by improving the memory update mechanism. Second, we plan to explore more sophisticated time-encoding methods to transform timestamps into features, instead of the current standard time encoding function. Third, we plan to build a paper recommendation web portal in a real-world academic setting and gather user feedback for iterative improvements.

\begin{acknowledgments}
This work is partially sponsored by NASA 80NSSC22K0144.\end{acknowledgments}

\bibliography{references.bib}

\appendix

\end{document}